\documentclass{article}
\usepackage{graphicx,spconf,amsmath,epsfig}
\usepackage{hyperref}
\usepackage{svg}
\usepackage{multirow}
\usepackage{fancyhdr}

\title{IONOSPHERIC SCINTILLATION FORECASTING USING MACHINE LEARNING}
\name{Sultan Halawa, Maryam Alansaari, Maryam Sharif, Amel Alhammadi, Ilias Fernini}
\address{Sharjah Academy for Astronomy Space Sciences and Technology \\University of Sharjah, Sharjah, United Arabic Emirates}
\begin{document}
%
\maketitle
\thispagestyle{fancy}

\begin{abstract}
This study explores the use of historical data from Global Navigation Satellite System (GNSS) scintillation monitoring receivers to predict the severity of amplitude scintillation, a phenomenon where electron density irregularities in the ionosphere cause fluctuations in GNSS signal power. These fluctuations can be measured using the S4 index, but real-time data is not always available. The research focuses on developing a machine learning (ML) model that can forecast the intensity of amplitude scintillation, categorizing it into low, medium, or high severity levels based on various time and space-related factors. Among six different ML models tested, the XGBoost model emerged as the most effective, demonstrating a remarkable 77\% prediction accuracy when trained with a balanced dataset. This work underscores the effectiveness of machine learning in enhancing the reliability and performance of GNSS signals and navigation systems by accurately predicting amplitude scintillation severity. 
\end{abstract}
\section{Introduction}
\label{sec:intro}

Ionospheric scintillation in Global Navigation Satellite System (GNSS) signals is fluctuations in the ionospheric electron density. Such variation in the ionosphere refractive index results in rapid fluctuations in the amplitude and phase of GNSS signals as they travel through the ionosphere \cite{crane1977ionospheric}. The transit of GNSS signals in the ionosphere creates deviations in this layer's refractive index, causing remarkable changes in the amplitude and phase of the signals \cite{crane1977ionospheric}. The scintillation in amplitude is evident in significant changes in the signal noise ratio (SNR) of the GNSS signals, which, in turn, affects the GNSS system performance stability, reliability, and accuracy. In addition, amplitude scintillation is a serious menace to the quality of services provided by the GNSS satellites, often negatively affecting the signal quality by the final users. Therefore, the primary means of quantifying the amplitude scintillation, especially in the evaluation of GNSS signals, is by using the S4 index. It measures the severity of the amplitude fluctuations very accurately. The S4 Index is a quantitative measure derived by dividing the standard deviation of the signal power by the mean signal power \cite{rino2011theory}. Not only does it provide a means of measure, but it is also widely used in operational applications and academic research. S4 index data is produced continuously by specialized scintillation monitoring receivers that constantly track the SNR of GNSS signals. This, in turn, provides continual access to real-time S4 index data, which allows fast identification and investigation of scintillation events \cite{yeh2019superposition}.  
    
In regions lacking a Satellite-Based Augmentation System (SBAS) or real-time ground correction, accurate ionospheric scintillation forecasting is crucial due to reliance on model-based GNSS corrections. Limited regional data availability can lead to inaccuracies in these models, compounded by the dynamic and nonlinear nature of ionospheric behavior influenced by solar and geomagnetic activities. To address this, advanced forecasting models are necessary to improve prediction accuracy, especially in data-scarce areas. This research aims to develop and recommend predictive models for ionospheric scintillation, contributing to better precision and reliability of satellite-based systems in regions without real-time correction systems.

Three major recent advances significantly contribute to machine learning for ionospheric scintillation. The work in paper \cite{darya2022amplitude} explores the potential of historical data from a single G.P.S. scintillation monitoring receiver to train various machine learning models to predict the intensity of amplitude scintillation. An evaluative comparison was made among six distinct ML models: decision trees, naıve Bayes, support vector machines (SVM), k-nearest neighbors (K.N.N.), boosted trees, and bagged trees. The bagged trees model emerged as the most effective, attaining 81\% prediction accuracy on a balanced dataset and 97\% on an imbalanced one. This study serves as a foundational reference for our project, particularly in the methodology.In contrast to [4], our research is geographically centered in Sharjah, U.A.E.. It broadens its data range by including four satellite constellations, which are GPS, GLONASS, GALILEO, and BEIDOU, instead of relying solely on GPS data. We further advanced our analysis by integrating different ML algorithms such as CatBoost, LightGBM, and XGBoost. Moreover, a vital methodological improvement in our study is using complete signal coordinates, moving away from the limited use of ionospheric pierce points used in \cite{darya2022amplitude}. This approach enhances the accuracy and relevance of amplitude scintillation forecasts for the region.

Building upon the methodologies outlined in \cite{darya2022amplitude}, a related study conducted in Brazil \cite{carvalho2022nowcasting} focuses on developing short-term predictive models for amplitude ionospheric scintillation using machine learning techniques, an aspect crucially considered in our paper for model evaluation. The research aimed to forecast the scintillation index S4 30 minutes in advance, employing a dataset sourced from a NASA-maintained database that includes detailed information on geomagnetic, solar, and interplanetary activities crucial for understanding the ionospheric conditions. The predictive analysis utilized three algorithms: Random Forest (RF), Artificial Neural Network (ANN), and Extreme Gradient Boosting (XGBoost). Among these, RF models emerged with better performances in the test set than ANN and XGB models. Notably, the least effective model still achieved a coefficient of determination of 0.87, indicating the robustness of the chosen methods. The study's results validated the usefulness of the selected dataset and the feature selection approach during the model development phase, leading to more effective models, as evidenced by several statistical tests. This research's comprehensive approach to dataset utilization and algorithm selection provided valuable insights. It served as a significant reference for our study in adopting machine learning models for forecasting ionospheric scintillation. 

Finally, the work in \cite{dey2021automatic} utilizes the XGBoost algorithm, a scalable machine learning classifier technique, to classify and detect ionospheric amplitude scintillation. This approach was applied to a comprehensive dataset from the Sao Jose station in Brazil, containing GPS data from 2012 to 2016 and 2018 to 2019. The training set comprises 243,745 instances, including 16,952 scintillation events detected between 2012-2016 and 2018. These samples constitute 95\% of the overall collected data. However, for the testing set aimed to evaluate and validate the performance model, 5\% of the trained data, in addition to untrained data collected in 2019, compressing 26,592 instances with 743 labeled as scintillation events. XGBoost method demonstrates improved accuracy compared to other created machine learning techniques: Neural Networks (NN), SVM, Decision Trees (DT), and Logistic Regression in modeling and predicting ionospheric disturbances. It handles data irregularities and reduces computational demands, achieving a high prediction accuracy of 99.88\%. This paper provided valuable insights into the model selection and management of large datasets, significantly contributing to the refinement of machine learning models in our large-scale, data-focused research.

The paper is divided into several subsections: The dataset used in this work is described in Section 2, followed by the methodology described in Section 3, results and discussion in Section 4, and finally conclusion in Section 5.

\section{Dataset}
\label{sec:Dataset}

\par In this work, we utilized observations of amplitude scintillation which is observed through S4 index. the total S4 index ($S4_{total}$) is defined as the standard deviation of the 50-Hz signal intensity(SI) normalized to the average SI over 60 seconds. 
\begin{equation} \label{eq:0}
    S4_{total} = \sqrt{\frac{(\text{SI}^2) - (\text{SI}^2)}{(\text{SI}^2)} - \frac{100}{\text{SNR}}\left[1 + \frac{500}{19 \text{SNR}}\right]},
  \end{equation}
  where SNR is the signal-to-noise ratia of the signal.
  
The index is obtained from a multi-constellation, multi-frequency GNSS receiver with scintillation monitoring capabilities at the Sharjah GNSS Station at SAASST (25.28, 55.46) in the United Arab Emirates. The data spans from Oct. 9, 2018, to Aug. 11, 2023, and it is retrieved from the Ionospheric Scintillation Monitoring Records (ISMR) files, with 1-second granularity. The dataset underwent a series of processing steps, as outlined below:

\begin{enumerate}
  \item Location and Multi-Path Interference Mitigation: The GNSS receiver is strategically positioned to minimize multi-path interference, with a 20-degree elevation threshold applied to filter out observations susceptible to such interference.
  \item Satellite Selection: Observations were narrowed to include only GPS, GLONASS, GALILEO, and BEIDOU satellites, excluding others satellites using satellite pseudorandom noise (PRN) code. Including  satellite PRN is important because its been observed that different satellites or constellations have different levels of ionospheric scintillation.\cite{9990125}
  \item S4 Correction: The S4 values underwent correction using the formula in (\ref{eq:1}) and (\ref{eq:2}):
  \begin{equation} \label{eq:1}
    X = S4^2_{total} - S4^2_{noise},
  \end{equation}
  \begin{equation} \label{eq:2}
    S4_{corrected}=\begin{cases}
    \sqrt{X}, & \text{if $X<0$}.\\    0, & \text{otherwise},
  \end{cases}
  \end{equation}
  \item Conversion to UTC: GPS week number (WN) and time of week (TOW) were converted to UTC values using the formula in (\ref{eq:3}):
  \begin{equation} \label{eq:3}
  \begin{split}
    \text{UTC}\textsc{time} = \text{timestamp(1980, 1, 6)} + \text{WN} \\ + \text{TOW} - \text{leapseconds}
  \end{split}
  \end{equation}
  \item Temporal Smoothing: A 5-minute average was applied to the elevation, azimuth, and S4 Corrected features for each Space Vehicle ID (SVID) to ensure the authenticity of S4 effects and eliminate noise.
  \item Solar Indices Integration: Three widely used solar indices (daily average Planetary Kp-index (KP), Sunspot Number (SSN), and F10.7 index) were incorporated into the dataset. These indices were sourced from \url{https://omniweb.gsfc.nasa.gov/form/dx1.html}.
  \item Data Classification: The data was categorized into three classes based on S4 values: Class 1 (S4 $<$ 0.2), Class 2 (0.2 $\leq$ S4 $<$ 0.3), and Class 3 (S4 $\geq$ 0.3). Class sizes were imbalanced, with the number of observations for classes 1, 2, and 3 being 12,309,632, 19,331, and 9,114, respectively, totaling 12,338,077 observations.
  \item Balanced Dataset Creation: To address the class imbalance, a balanced dataset was generated by randomly sampling 9,000 observations from each class, resulting in a total of 27,000 observations.
\end{enumerate}

\section{Methodology}
\label{sec:Methodology}

The machine learning models employed in this work encompass various algorithms. The following paragraphs provide an overview of the models utilized:

\textbf{K-Nearest Neighbors (KNN):} The KNN algorithm works by finding data close to other data points with similar features or meanings. Therefore, the main concept of the KNN is to designate a category for the data point based on the similarity found among its closest neighbors. In other words, the algorithm assigns a category to a data point based on its similarities to its nearest neighbors \cite{liao2002use}.

\textbf{Support Vector Machine (SVM):} SVM is adept at constructing an optimal decision boundary, effectively categorizing data into their designated classes. This boundary, commonly referred to as the hyper-plane, is pivotal role in classification. The data points that lie nearest to this hyper-plane and significantly influence its placement are identified as support vectors; hence the name allocated \cite{steinwart2008support}.

\textbf{Naive Bayes (NB):} NB classifiers operate under the assumption that attribute values are conditionally independent of each other given the target value or class \cite{chen2020novel}. Therefore, in our context, each feature independently predicts an S4 value without relying on inter-feature dependencies.

\textbf{CatBoost:} CatBoost is part of the Gradient Boosted Decision Trees framework and is particularly useful for dealing with diverse and categorical data types in machine learning \cite{hancock2020catboost}. It uses decision trees for regression and classification tasks and has two key features: the ability to handle categorical data directly and the application of gradient-boosting techniques. Unlike other decision tree-based methods, CatBoost does not require pre-processing steps and can efficiently handle a mix of categorical and numerical variables. It also uses innovative methods such as "ordered encoding" to enhance the encoding process.

\textbf{LightGBM:} LightGBM is an efficient and powerful gradient-boosting tool that uses decision trees to increase model efficiency. Its design is centered around a histogram-based algorithm that categorizes continuous feature values into distinct bins, leading to faster training. This method reduces memory usage by converting continuous values to discrete bins, resulting in significant savings. LightGBM's speed and efficiency make it an excellent option for complex machine-learning tasks requiring agility and resource optimization \cite{mccarty2020evaluation}.

\textbf{XGBoost:} XGBoost is an optimized gradient boosting library widely used for its efficiency and predictive accuracy. It implements a regularized boosting algorithm and is effective across various machine-learning tasks \cite{chen2016xgboost}.

Following the data preprocessing steps in Section \ref{sec:Dataset}, which included handling missing values and ensuring data completeness. The next stage, importing the dataset into our analysis environment. Subsequently, we defined the features and target variables. Employing the holdout technique, the dataset was split into training and testing sets, allocating 80\% for training and reserving 20\% for testing. With the data appropriately partitioned, the machine learning models were fitted using the training dataset. Each model underwent a training process to learn the patterns within the data. Upon completion of the training phase, predictions were generated using the testing dataset. The predictive performance of each model was then evaluated by comparing the predicted values against the actual values. Key evaluation metrics were calculated to assess the models' effectiveness in capturing and classifying instances, including accuracy, precision, recall.

Subsequently, we employed an exhaustive search method using the GridSearchCV function. It systematically explores a specified hyper-parameter grid to find the combination that yields the best performance for a given model.

\section{Results and discussion}
\label{sec:Results and discussion}

The performance of the different ML models used in this work was tested regarding their accuracy, precision, and recall \cite{davis2006relationship}. Accuracy, in this case, is defined as the number of correct predictions divided by the total number of predictions, as represented in equation (\ref{eq:4}): 

\begin{equation} \label{eq:4}
    \text{Accuracy} = \frac{\text{TP} + \text{TN}}{\text{TP} + \text{TN} + \text{FP} + \text{FN}} ,
\end{equation}

where TP, TN, FP, and FN are the number of true positives, true negatives, false positives, and false negatives, respectively. Furthermore, precision and recall can be defined in equation (\ref{eq:5}) and (\ref{eq:6}).
\begin{equation} \label{eq:5}
    \text{Precision} = \frac{\text{TP}}{\text{TP} + \text{FP}} ,
\end{equation}
\begin{equation} \label{eq:6}
    \text{Recall} = \frac{\text{TP}}{\text{TP} + \text{FN}} .
\end{equation}

\begin{table}[]
\centering
\begin{tabular}{|c|l|c|}
\hline
\textbf{Model} & \multicolumn{1}{c|}{\textbf{Parameters}}                                                                                                        & \textbf{\begin{tabular}[c]{@{}c@{}}Balanced \\ Dataset \\ Accuracy\end{tabular}} \\ \hline
KNN            & \begin{tabular}[c]{@{}l@{}}Leaf size = 1 \\ Number of neighbors = 12 \\ P = 1\end{tabular}                                                      & 65.7\%                                                                          \\ \hline
SVM            & \begin{tabular}[c]{@{}l@{}}C = 1000 \\ Gamma = 1 \\ Kernel function = RBF\end{tabular}                                                          & 65.3\%                                                                          \\ \hline
\begin{tabular}[c]{@{}c@{}}Naive \\ Bayes\end{tabular}    & Kernel type = Gaussian                                                                                                                          & 54.8\%                                                                          \\ \hline
CatBoost       & \begin{tabular}[c]{@{}l@{}}Classes Count = 3 \\ Learning Rate = 0.1                            \\ L2 Leaf Reg = 3 \\ Max Depth = 6\end{tabular} & 75.3\%                                                                          \\ \hline
LightGBM       & \begin{tabular}[c]{@{}l@{}}Objective = multiclass \\ Number of Classes = 3\end{tabular}                                                         & 74.5\%                                                                          \\ \hline
XGBoost        & \begin{tabular}[c]{@{}l@{}}Number of Classes = 3 \\ Learning Rate = 0.3 \\ Max Depth = 9\end{tabular}                                           & 76.7\%                                                                          \\ \hline
\end{tabular}
\caption{Results of Machine Learning Models}
\label{table:1}
\end{table}

The performance and parameters of the developed predictive models for ionospheric scintillation forecasting are presented in Table \ref{table:1}. XGBoost demonstrates the highest accuracy at 76.7\%, closely followed by CatBoost and LightGBM. While KNN and SVM exhibit reasonable accuracy, NB show slightly lower performance in this context. We note that the baseline accuracy of the balanced dataset is 33.3\% since all categories of the balanced dataset have an equal number of observations. A simple model that randomly allocates S4 values to either of the three categories can achieve the baseline accuracy. As a result, the model's accuracy improvement over the baseline is considerably higher, at about 43.4\%.

\begin{table}[]
\centering
\begin{tabular}{|cc|ccc|c|}
\hline
\multicolumn{2}{|c|}{\multirow{2}{*}{}}                   & \multicolumn{3}{c|}{\textbf{Ground Truth}}                            &                                                            \\ \cline{3-6} 
\multicolumn{2}{|c|}{}                                    & \multicolumn{1}{c|}{Class 1} & \multicolumn{1}{c|}{Class 2} & Class 3 & Precision                                                  \\ \hline
\multicolumn{1}{|c|}{\multirow{3}{*}{\rotatebox[origin=c]{90}{ Predict.}}} & Class 1 & \multicolumn{1}{c|}{1498}    & \multicolumn{1}{c|}{75}      & 219     & 81.4\%                                                     \\ \cline{2-6} 
\multicolumn{1}{|c|}{}                          & Class 2 & \multicolumn{1}{c|}{66}      & \multicolumn{1}{c|}{1456}    & 272     & 77.4\%                                                     \\ \cline{2-6} 
\multicolumn{1}{|c|}{}                          & Class 3 & \multicolumn{1}{c|}{279}     & \multicolumn{1}{c|}{351}     & 1187    & 70.7\%                                                     \\ \hline
\multicolumn{1}{|c|}{}                          & Recall  & \multicolumn{1}{c|}{83.6\%}  & \multicolumn{1}{c|}{81.2\%}  & 65.4\%  & \begin{tabular}[c]{@{}c@{}}Accuracy \\ 76.7\%\end{tabular} \\ \hline
\end{tabular}
\caption{Confusion Matrix for Dataset Using XGB.}
\label{table:2}
\end{table}

In Table \ref{table:2}, the matrix highlights the model's ability to make accurate predictions within each class, showcasing notable precision percentages, such as 81.4\% for low severity, 77.4\% for mid severity, and 70.7\% for high severity. Additionally, the recall values demonstrate the model's sensitivity to each severity class, with percentages of 83.6\%, 81.2\%, and 65.4\% for low, mid, and high severity, respectively. These metrics collectively contribute to an overall accuracy of 76.7\%. The discussion emphasizes the balanced performance of the XGBoost model, offering a comprehensive evaluation of its strengths and areas for improvement in predicting amplitude scintillation severity.

While our study on ionospheric scintillation forecasting offers valuable insights using data from a single GNSS receiver at the Sharjah GNSS Station, it's crucial to acknowledge certain limitations. The U.A.E.'s relatively small geographical size minimizes the impact of regional variations on ionospheric behavior, suggesting that our findings may be applicable across the country. However, the need for broader data sources persists, emphasizing the importance of incorporating information from multiple GNSS receivers spread across different municipalities. Additionally, the reliance on historical data, without real-time integration, restricts our ability to provide instant forecasts, a limitation that can impact applications requiring immediate decision-making. Moreover, our study advocates for establishing a centralized GNSS receiver system in the U.A.E. to streamline data aggregation, ensure standardized monitoring, and enhance the robustness of nationwide scintillation forecasting models. These limitations signal opportunities for future research to expand datasets and real-time integration to refine ionospheric scintillation predictions in the U.A.E further.

\section{Conclusion}
\label{sec:Conclusion}

This work focused on predicting the severity of amplitude scintillation, which varies spatially and temporally, utilizing ML techniques. Six distinct machine learning models were evaluated, among which XGBoost emerged as the most effective, achieving an accuracy of 76.7\%. Future plans include broadening the scope of this research by applying these models to data collected from various stations across different regions. Additionally, we intend to explore the potential of deep learning models, which might be particularly advantageous given the extensive volume of training data available. Our findings support the idea that there is a need for a centralized GNSS receiver system. Such a system would significantly enhance the efficiency of data collection, standardization, and monitoring processes, highlighting the essential role of artificial intelligence in optimizing these critical infrastructures.

\vfill
\pagebreak

\bibliographystyle{IEEEbib}
\bibliography{main.bib}

\begin{thebibliography}{10}

\bibitem{crane1977ionospheric}
Robert~K Crane,
\newblock ``Ionospheric scintillation,''
\newblock {\em Proceedings of the IEEE}, vol. 65, no. 2, pp. 180--199, 1977.

\bibitem{rino2011theory}
Charles Rino,
\newblock {\em The theory of scintillation with applications in remote sensing},
\newblock John Wiley \& Sons, 2011.

\bibitem{yeh2019superposition}
Wen-Hao Yeh, Chi-Yen Lin, Jann-Yenq Liu, Shih-Ping Chen, Tung-Yuan Hsiao, and Cheng-Yung Huang,
\newblock ``Superposition property of the ionospheric scintillation s4 index,''
\newblock {\em IEEE Geoscience and Remote Sensing Letters}, vol. 17, no. 4, pp. 597--600, 2019.

\bibitem{darya2022amplitude}
Abdollah~Masoud Darya, Aisha~Abdulla Al-Owais, Muhammad~Mubasshir Shaikh, and Ilias Fernini,
\newblock ``Amplitude scintillation forecasting using bagged trees,''
\newblock in {\em IGARSS 2022-2022 IEEE International Geoscience and Remote Sensing Symposium}. IEEE, 2022, pp. 2275--2278.

\bibitem{carvalho2022nowcasting}
Ot{\'a}vio Carvalho, Ricardo Yvan de La~Cruz Cueva, Alex~Oliveira Barradas~Filho, et~al.,
\newblock ``Nowcasting of amplitude ionospheric scintillation based on machine learning techniques,''
\newblock {\em IEEE Transactions on Aerospace and Electronic Systems}, vol. 58, no. 6, pp. 4917--4927, 2022.

\bibitem{dey2021automatic}
Abhijit Dey, Manhal Rahman, Devanaboyina~Venkata Ratnam, and Nitin Sharma,
\newblock ``Automatic detection of gnss ionospheric scintillation based on extreme gradient boosting technique,''
\newblock {\em IEEE Geoscience and Remote Sensing Letters}, vol. 19, pp. 1--5, 2021.

\bibitem{9990125}
Abdollah~Masoud Darya, Muhammad~Mubasshir Shaikh, Ilias Fernini, and Hamid AlNaimiy,
\newblock ``Abnormal phase scintillation observed from glonass signals,''
\newblock in {\em 2022 International Conference on Electrical and Computing Technologies and Applications (ICECTA)}, 2022, pp. 89--92.

\bibitem{liao2002use}
Yihua Liao and V~Rao Vemuri,
\newblock ``Use of k-nearest neighbor classifier for intrusion detection,''
\newblock {\em Computers \& security}, vol. 21, no. 5, pp. 439--448, 2002.

\bibitem{steinwart2008support}
Ingo Steinwart and Andreas Christmann,
\newblock {\em Support vector machines},
\newblock Springer Science \& Business Media, 2008.

\bibitem{chen2020novel}
Shenglei Chen, Geoffrey~I Webb, Linyuan Liu, and Xin Ma,
\newblock ``A novel selective na{\"\i}ve bayes algorithm,''
\newblock {\em Knowledge-Based Systems}, vol. 192, pp. 105361, 2020.

\bibitem{hancock2020catboost}
John~T Hancock and Taghi~M Khoshgoftaar,
\newblock ``Catboost for big data: an interdisciplinary review,''
\newblock {\em Journal of big data}, vol. 7, no. 1, pp. 1--45, 2020.

\bibitem{mccarty2020evaluation}
Dakota~Aaron McCarty, Hyun~Woo Kim, and Hye~Kyung Lee,
\newblock ``Evaluation of light gradient boosted machine learning technique in large scale land use and land cover classification,''
\newblock {\em Environments}, vol. 7, no. 10, pp. 84, 2020.

\bibitem{chen2016xgboost}
Tianqi Chen and Carlos Guestrin,
\newblock ``Xgboost: A scalable tree boosting system,''
\newblock in {\em Proceedings of the 22nd acm sigkdd international conference on knowledge discovery and data mining}, 2016, pp. 785--794.

\bibitem{davis2006relationship}
Jesse Davis and Mark Goadrich,
\newblock ``The relationship between precision-recall and roc curves,''
\newblock in {\em Proceedings of the 23rd international conference on Machine learning}, 2006, pp. 233--240.

\end{thebibliography}

\end{document}